\documentclass[twocolumn,showpacs,preprintnumbers,amsmath,amssymb]{revtex4}

\usepackage{graphicx}
\usepackage{dcolumn}
\usepackage{bm}

\newcommand{\be}{\begin{equation}}
\newcommand{\en}{\end{equation}}
\newcommand{\bea}{\begin{eqnarray}}
\newcommand{\ena}{\end{eqnarray}}
\newcommand{\Det}{\hbox{Det}}
\newcommand{\hbo}{\hbox to 1 true cm {\hfill } }
\newcommand{\tr}{\hbox{tr}}

\def\dslash{\partial\kern-.5em\slash}

\begin{document}


\title{ Confinement and the quark Fermi-surface in SU(2N) QCD-like theories }

\author{Kurt Langfeld$^a$, Bjoern H.~Wellegehausen$^b$, Andreas Wipf$^b$}

\affiliation{%
$^a$School of Mathematics and Statistics, University of Plymouth \\
Plymouth PL4 8AA, UK \\
$^b$TPI, Friedrich-Schiller-Universit\"at Jena \\
D-07743 Jena, Germany
}%

\date{\today}

\begin{abstract}
Yang-Mills theories with a gauge group $SU(N_c\not=3)$ and quark matter in
the fundamental representation share many properties with the theory
of strong interactions, QCD with $N_c=3$. We show that, for $N_c$ even and
in the confinement phase, the gluonic average of the 
quark determinant is independent
of the boundary conditions, periodic or anti-periodic ones. We then argue
that a Fermi sphere of quarks can only exist under extreme conditions when
the centre symmetry is spontaneously broken and colour is liberated. Our
findings are supported by lattice gauge simulations for $N_c=2 \ldots 5$ and
are illustrated by means of a simple quark model.
\end{abstract}

\maketitle
In particle physics, understanding the theory of strong interactions, i.e.,
QCD, under extreme conditions is key for a successful description of
exotic matter which features in the evolution of the universe
or cold compact stars. The property of QCD which is most relevant for
shaping the QCD phase diagram (if presented as a function of temperature and
matter density) is colour confinement. Under normal conditions,
confinement implies that quark matter is organised in terms of
hadrons only, and quarks and gluons are only part of the particle
spectrum under extreme conditions.

By means of lattice gauge simulations, a precise picture of hot QCD
matter at small densities has emerged over the last two decades:
Centre symmetry is spontaneously broken in the hot de-confinement phase
for temperatures above a certain critical value. Colour
is liberated and, in a theory with heavy quarks only, the Polyakov
line expectation value serves as an order
parameter~\cite{Svetitsky:1982gs,Svetitsky:1985ye}.

On the other hand, the situation at high densities and small temperatures
is far
from being clear. The reason is the lack of first principle QCD results
which would help to scrutinise the proposals for the properties of
matter in this regime. Lattice gauge simulation techniques cannot be applied
because of the severity of the so-called sign or overlap problem.
These problems are absent for the so-called two-colour QCD, and intriguing
results, even for large quark chemical potential, have been accumulated
over the recent
years~\cite{Kogut:2001na,Conradi:2007kr,Hands:2005yq}.
For an investigation of cold and dense matter in $SU(N_c \ge 3)$
QCD(-like) theories, we are still awaiting major conceptual achievements.
Promising recent attempts abandon standard lattice Monte-Carlo techniques
and are based upon stochastic quantisation~\cite{Aarts:2008rr} or
worldline numerics~\cite{Dunne:2009zz}.

Perturbative QCD and QCD inspired quark models have been a valuable tool
for revealing mechanisms which might operate in the cold and dense phase
of QCD. On this basis, the QCD phase diagram has gained a lot of renewed
interest when findings suggested that its structure is far more complex
than it had been suggested for decades: at the highest densities, it is
expected that quark matter is forming a colour-superconductor which
carries along a rich phase structure on its own (for a recent review
see~\cite{Alford:2007xm}).
Studies of the Gross-Neveu model indicate that
dense QCD (at low temperatures) might form an inhomogeneous
baryonic crystal~\cite{Schnetz:2004vr,Schnetz:2005ih,Boehmer:2007ea}.

Employing arguments based upon the large $N_c$ expansion, it has been
recently suggested that the tight relation between confinement and
spontaneous chiral symmetry breaking (inherent for zero density QCD)
gets alleviated~\cite{McLerran:2007qj,McLerran:2008ua}.
In the phase diagram as a function of chemical
potential and temperature, the phase boundary for chiral restoration
might deviate from the boundary for deconfinement. Most interesting,
a phase for which confinement is still intact while chiral symmetry is
restored has attracted a lot of interest. This so-called {\it quarkyonic}
phase is characterised by a Fermi sphere of quarks while the outer shell
of the Fermi sphere necessarily  consists of baryons because of
confinement~\cite{McLerran:2007qj}.

In this paper, we point out that the properties of dense quark matter
in $SU(N_c)$ QCD-like theories are vastly different depending on whether
the number of colours, $N_c$, is even or odd. In the confining phase,
the quark determinant is averaged over centre-transformed gluonic
background fields. Underpinned by lattice gauge
simulations, we will show that this averaged quark determinant
is insensitive to the boundary conditions of the quarks, periodic or
anti-periodic ones. Since the formation of a quark Fermi sphere is crucially
linked to anti-periodic boundary conditions (see below for an illustration
by a quark model), we will argue that
a Fermi sphere of quarks can only exist in the deconfined phase
of $SU(N_c)$ QCD-like theories with $N_c$ being even.

QCD-like theories are $SU(N_c)$ Yang-Mills theories coupled to fermions
(``quarks'') in the fundamental representation. We here adopt the lattice
regularisation based upon a toroidal space-time lattice with lattice spacing
$a$ and extension $N_t \times N_s^3$. The gluonic degrees of freedom
$U_\mu (x) \in SU(N_c)$ satisfy periodic boundary conditions, e.g.,
$
U_\mu (x_0+N_t a, \vec{x}) \; = \; U_\mu (x_0, \vec{x}) \; .
$
Quark fields $q(x)$ are associated with the lattice sites. Because of
the Fermi statistics of the bare quark fields, these fields satisfy
antiperiodic boundary conditions:
$$
q(x_0 + N_t a ,\vec{x}) \, = \, (-1) \, q(x_0,\vec{x}).
$$
Using the Wilson action for the gluonic fields, the partition function
is given by
\bea
{\cal Z } &=& \int {\cal D}U_\mu \; \Det _{A} M[U] \;
\mathrm{e}^{
\frac{\beta}{N_c} \sum _{x, \mu \nu } \, \mathrm{Re} \, \tr \, P_{\mu \nu}(x)} ,
\label{eq:1}
\ena
where $\beta = 2N_c/g^2$ is given in terms of the Yang-Mills gauge coupling
$g$, and $P_{\mu \nu }$ is the standard plaquette.
The determinant in (\ref{eq:1}) arises from the integration over the quark
fields. The subscript ``A'' indicates that the quarks were subjected to
antiperiodic boundary conditions. Here, we work with Wilson
quarks where
\bea
M[U] &=& (m + 4) \delta _{xy} - \frac{1}{2} \sum _{\mu =1}^4
\Bigl[ (1 - \gamma _\mu ) \, U_\mu(x)
\, \delta _{x + \mu, y}
\nonumber \\
&+&  (1 + \gamma _\mu ) \, U^\dagger _\mu (x-\mu)
\, \delta _{x -\mu, y} \, \Bigr] \; ,
\label{eq:3}
\ena
where $\gamma _\mu $ are the Hermitian Dirac matrices and $m$ is
the current quark mass in units of the lattice spacing.
For an investigation of the quark determinant, we insert
$$
1 \, = \, \int dQ \; \delta \Bigl( Q - \Det _{A} M[U] \Bigr)
$$
into (\ref{eq:1}) to write the partition function as
\bea
{\cal Z } &=& \int dQ \, Q \, P_A(Q) \, ,
\label{eq:4} \\
P_A(Q) &=& \int {\cal D}U_\mu \,\delta \Bigl( Q - \Det _{A} M[U] \Bigr)
\, \mathrm{e}^{S_{\rm YM}[U]} ,
\label{eq:5}
\ena
where $P_A(Q)$ is the probability distribution of the quark determinant.
Note that $P_A(Q)$ can be calculated using Monte-Carlo techniques for
{\it pure} Yang-Mills theory. Note that the probability distribution
of the quark determinant of full QCD, $P_\mathrm{full}(Q)$, is related to
that of pure Yang-Mills theory, i.e., $P_A(Q)$, by $P_\mathrm{full}(Q)
= Q P_A(Q)$. In particular for large lattices, little statistics is
expected for the large $Q$ regime which is more relevant
for the simulation of Yang-Mills theory with
dynamical quarks included. If this regime is under considerations,
it is advisable to include the determinant in the
simulation~\cite{Duane:1987de} and to use refined simulation techniques
such as those in~\cite{Fodor:2001pe,Fodor:2004nz} when finite temperatures
and densities are addressed. The intermediate $Q$ regime will turn out
to be  sufficient to illustrate our findings below, and thus
only quenched simulations are used throughout this paper.

Let us now consider a Roberge-Weiss transformation~\cite{Roberge:1986mm}
in the gluonic functional integral in (\ref{eq:5}).
For a fixed $x_0=t$, we consider
\bea
U_0 (t,\vec{x}) & \rightarrow & z_n \, U_0 (t,\vec{x}) \qquad
\forall \vec{x}\,,
\label{eq:6} \\
z_n &=& \exp \Bigl\{ \frac{2\pi i}{N_c} n \Bigr\},\quad
0 \le n \le N_c-1 .
\label{eq:7}
\ena
Given the invariance of the gluonic action and of the Haar measure
${\cal D}U_\mu $, we find:
\be
P_A(Q) = \int {\cal D}U_\mu \,\delta \Bigl( Q - \Det _{A} M[z_nU] \Bigr)
\, \mathrm{e}^{S_{\rm YM}[U]} .
\label{eq:8}
\en
Reintroducing the quark fields for a moment,
\be
\Det _{A} M[z_nU] = \int {\cal D}q \, {\cal D}\bar{q} \,
\exp \Big\{ \sum _{xy} \bar{q}(x) M_{xy} q(y) \Big\} .
\label{eq:9}
\en
we explore the virtue of the transformation:
\bea
U_\mu ^\Omega (x) &=& \Omega (x) \, U_\mu (x) \, \Omega ^\dagger (x+\mu) ,
\label{eq:15} \\
q^\Omega (x) &=& \Omega (x) \, q(x) \; ,
\label{eq:16} \\
\Omega (x_0,\vec{x} ) &=& z_n \; \hbox{for} \; x_0>t \; , \; \; \;
\Omega (x_0,\vec{x} ) = 1 \; \hbox{else}.
\label{eq:17}
\ena
Note that $t$ has been defined above (\ref{eq:6}).
We point out that $\Omega (x)$ satisfies the boundary condition
$
\Omega (x_0 + N_t a,\vec{x} ) = z_n \Omega (x_0,\vec{x} ) .
$
\begin{figure*}
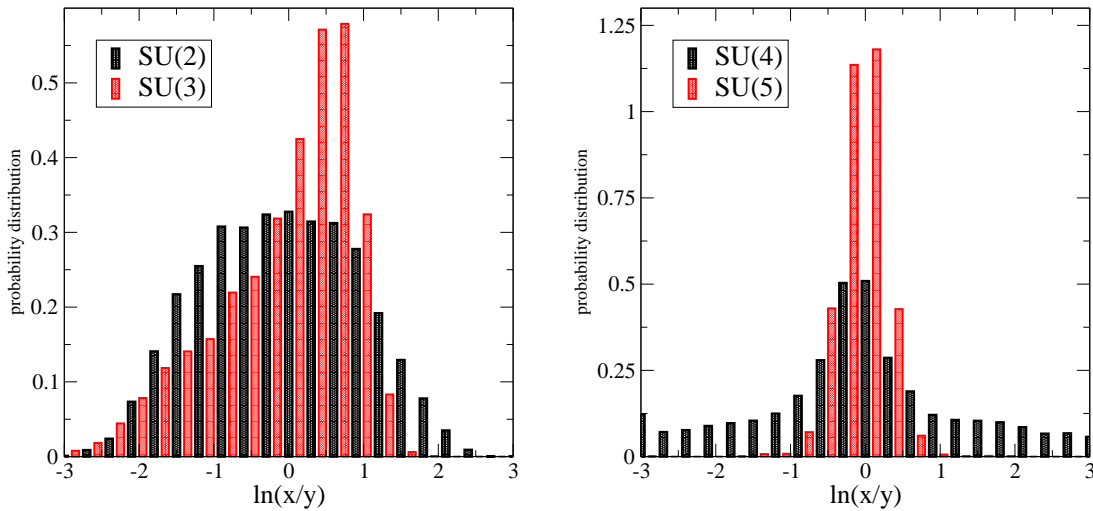

\includegraphics[height=6.7cm]{histo_23.eps} \hspace{0.5cm}
\includegraphics[height=6.7cm]{histo_45.eps}
\caption{\label{fig:1} Probability distribution for $\ln r$
in comparison for SU(2) and SU(3) (left) and for SU(4) and SU(5) (right).
}
\end{figure*}
The transformation (\ref{eq:15}) does not change the boundary conditions
of the link fields and leaves the Yang-Mills action invariant. It can be
therefore considered as a gauge transformation of pure $SU(N_c)$ Yang-Mills
theory. Note, however, that the transformation (\ref{eq:16}) does change
the boundary conditions of the quark fields:
\be
q^\Omega (x_0 + N_ta, \vec{x}) \; = \; (-1) \, z_n \, q^\Omega
(x_0,\vec{x}) .
\label{eq:18}
\en
It therefore does not qualify as a gauge transformation of the full theory.
It, however, reveals an important property of the distribution
$P_A(Q)$. Specialising to
an {\it even } number of colours and choosing
$n =N_c/2$, $z_n = - 1$,
we obtain from (\ref{eq:9}) using (\ref{eq:15}-\ref{eq:17}):
\be
\Det _{A} M[z_nU] \; = \; \Det _{P} M[U] \,,
\label{eq:19}
\en
where the subscript ``P'' signals that the quark fields now obey
{\it periodic } boundary conditions. Inserting the last result (\ref{eq:19})
into (\ref{eq:8}) and using the gauge invariance of Yang-Mills action and
Haar measure, we finally obtain:
\bea
P_A(Q) &=& \int {\cal D}U_\mu \,\delta \Bigl( Q - \Det _{P} M[U] \Bigr)
\, \mathrm{e}^{S_{\rm YM}[U]}
= P_P(Q).
\nonumber
\ena
In QCD-like theories for an even number of colours and, at least, for a
finite volume (see comment below), the partition function is independent
of the choice of the quark boundary conditions, periodic or antiperiodic
ones. In the infinite volume limit, the centre symmetry corresponding to
the transformation (\ref{eq:7}) is not always realised: it has been known for
a long time~\cite{Svetitsky:1982gs,Svetitsky:1985ye} that this
symmetry is spontaneously broken under extreme conditions, high temperature
and/or fermion densities. While periodic boundary conditions are associated
with an instability of the partition function for vanishing temperature
corresponding to Bose-Einstein condensation, antiperiodic boundary
conditions are the essential ingredient for building up a Fermi sphere
in the dense phase. Hence, we argue that in the confining phase of
$SU(2N)$ QCD-like theories, a quark Fermi surface is unlikely to exist.
We stress, however, that the deconfinement phase at high densities (and low
temperatures) might well feature a Fermi sphere of quarks.

We are now going to illustrate our findings using
lattice gauge simulations as well as a simple quark model.
We have carried out simulations for SU($N_c$), $N_c=2,3,4,5$ using a
$4^4 $ space time lattice and a Dirac mass $m=0.01$.
The Wilson parameters $\beta $ have been chosen
such that in all cases the lattice spacing $a$ in units of
string tension $\sigma $
was roughly constant, i.e., $\sigma a^2 = 0.467(10)$. Keeping $\sigma a^2$
fixed when the number of colours $N_c$ is increased also complies with
the so-called 't~Hooft limit where $g^2 N_c = $constant. The simulation
parameters are summarised in table~\ref{tab:1}.

\begin{table}[h]
\begin{tabular}{ccccc}
group & \quad SU(2)\quad  & \quad SU(3) \quad & \quad SU(4) \quad &
\quad SU(5) \\
$\beta $ & $2.2$ & $ 5.63 $ & $10.99$ &  $16.57$ \\
$g^2_\mathrm{naive} N_c$ & $3.63$ & $3.2$ & $2.9$ & $ 3.0$
\end{tabular}
\caption{ \label{tab:1} Simulation parameters;
$9902$ configurations were used for SU(2) and SU(3), $8000$ for SU(4)
and $4500$ for SU(5) to estimate the expectation values. }
\end{table}

The determinants
have been calculated exactly using the standard LU-decomposition.
In order to explore the sensitivity of the quark determinants to
the boundary condition, we define
\be
x \; = \; \frac{\Det _P M[U]}{\Det _A M[U=1]} ,\quad
y \; = \; \frac{\Det _A M[U]}{\Det _A M[U=1]} ,
\label{eq:21}
\en
and consider the
ratio $r[U] := x/y $ for a given lattice configuration.
For $N_c$ {\it even}, we find with the results above that
\be
r \left[ z_n U\right] = 1/ r[U] \; \; \;
\Rightarrow \; \; \; \left\langle r \right\rangle
= \left\langle 1/r \right\rangle
\label{eq:22}
\en
if the centre-symmetry is realised. Our numerical findings for these
expectation values are summarised in table~\ref{tab:2}.

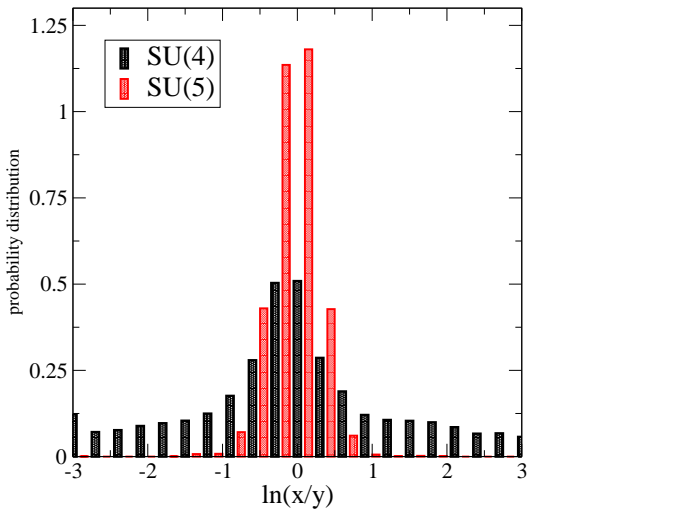
\begin{table}[h]
\begin{tabular}{ccccc}
 & \quad SU(2)\quad  & \quad SU(3) \quad & \quad SU(4) \quad &
\quad SU(5) \\
$\langle r \rangle $   & $1.66 \pm 0.02$ & $1.365 \pm 0.009$ &
$2.9977 \pm   0.07$ & $1.0469 \pm  0.005$ \\
$\langle 1/r \rangle $ & $1.65 \pm 0.02$ & $1.587 \pm   0.02$ &
$ 2.87 \pm 0.07 $ & $1.0571 \pm 0.007$ \\
\end{tabular}
\caption{ \label{tab:2}  Sensitivity parameters $\langle r \rangle $ and
$\langle 1/r \rangle $ for several gauge groups. }
\end{table}

We here find a clear coincidence between $\langle r \rangle $ and
$\langle 1/r \rangle $ for the gauge group SU(2) while the
corresponding parameters for SU(3) are significantly different.
We also observe a tendency that the difference fades away for increasing
number of colours $N_c$. We finally present the probability
distribution for the variable $\ln r$ in figure~\ref{fig:1}.
For gauge groups with $N_c$ even,
we expect that the corresponding histograms are symmetric with
respect to $\ln r  \to - \ln r $ (see~\ref{eq:22}). This expectation
is nicely confirmed for the gauge groups SU(2) and SU(4). A clear
asymmetry is observed for SU(3) while the asymmetry is very small
for SU(5) if present at all. We also empirically
observe that the width of the probability distribution decreases for
an increasing number of colours.

Let us finally illustrate by means of a quark model how centre sector
tunneling eliminates the quark Fermi surface.
The theory which we are
proposing is essentially a free theory of quarks which, however, interact
with a constant centre background field. We here work in the ab initio
continuum formulation. Parameterizing the centre background field by
($H = \mathrm{diag}(1,\ldots, 1, 1-N_c)/N_c$ from the Cartan algebra)
\be
A_n = 2\pi n \, T \, H ,  \quad 0 \le n \le N_c -1 ,
\label{eq:25}
\en
(where $T$ is the temperature)
the partition function of our model is in Euclidean space
\be
Z =  \sum _n p_n \, \frac{ \int {\cal D} q{\cal D} \bar{q}
\exp \{\bar{q} (i\dslash + (A_n + i\mu) \gamma_0 +  im ) q \} }{
 \int {\cal D} q{\cal D} \bar{q}
\exp \{\bar{q} (i\dslash +  im) q \} } ,
\label{eq:26}
\en
where $m$ is the quark mass and $\mu $ is quark chemical potential.
Thereby, $p_n$ is the probability that the centre sector $n$ is attained.
In the quenched approximation, all centre sectors occur with equal
probability, i.e., $p_n=1/N_c$. Note, however, that in a more QCD relevant 
setting dynamical quarks induce
a bias towards the trivial centre sector, i.e., $p_0 > p_{n\not=0}$. 
Furthermore, $A_n$ acts as
constant gauge field which can be eliminated from the action at the expense
of changing the boundary conditions for the quark fields. The partition
function can be calculated in closed form:
\bea
Z = \sum _n p_n
\prod _{\vec{p}, \vec{q}} \left( 1 + z_n \mathrm{e}^{-[E(\vec{p}) - \mu]/T }
\right)
\left( 1 + z_n^\dagger \mathrm{e}^{- [E(\vec{q}) + \mu]/T } \right) ,
\nonumber
\ena
where $z_n$ is related to the trace of the Polyakov line $P$ line by
$$ 
 P = \exp \{ i A_n /T \} , \hbo 
\frac{1}{N_c} \, \tr P = z_n 
$$
Moreover, the product extends over spatial momenta $\vec{p}$ and
$E (\vec{p}) =(m^2 + \vec{p})^ {1/2} $.
Expanding the brackets, because of the sum over the centre elements, 
only states with vanishing $N$-ality contribute to the partition function
asserting confinement. We here focus on the BEC instability.
We consider temperatures $T$ which are small compared to the mass gap
implying that we may neglect the contribution of antiquarks to the
free energy for $\mu \stackrel{>}{_\sim} m$. The free energy is
then given by $ \ln Z \approx \ln \sum_n p_n \, \rho_n $ with
\bea
\rho _n &=& \exp \Bigl\{ \frac{V}{\pi^2} \int _m^\infty dE \,
E \, \sqrt{E^2-m^2}
\ln \left( 1 + z_n \mathrm{e}^{- \frac{E - \mu}{T} }
\right) \Bigr\} ,
\nonumber
\ena
where $V$ is the spatial volume. Because of the spin of the quarks, we
have used two states per Matsubara mode.
For the search for Fermi surface effects, it is most instructive
to study the baryon number density:
\bea
b &=& \frac{T}{V} \,  \frac{\partial \ln Z}{\partial \mu }
= \frac{1}{\pi^2} \, \int _m^\infty dE \; E \; \sqrt{E^2-m^2} \; \rho
(E,T,\mu) ,
\nonumber
\ena
where $\rho $ is defined by
\bea
\rho (E,T,\mu) &=& \sum _n \frac{z_n}{ \mathrm{e}^{[E - \mu]/T }
+ z_n } \; w_n \; ,
\label{eq:32}
\ena
with the weights $w_n = p_n \, \rho_n / \sum_i p_i \, \rho _i$.
In the deconfined phase, tunneling between centre sectors stops because of
spontaneous symmetry breaking (on top of the explicit breaking
by the quark determinant).
The sum over the centre elements collapses to the trivial centre element
$p_0=1$, $p_{n\not=0}=0$ implying
$w_0=1$, $w_{n \not=0}=0$. In this case, $\rho (E,T,\mu) $ can be
interpreted as spectral density. This is given by the familiar
Fermi function
$
\rho _\mathrm{decon} (E,T,\mu) = [ \mathrm{e}^{[E - \mu]/T } + 1 ]^{-1} \; ,
$
which features a Fermi surface for $E \approx \mu $.

Let us now consider the confinement phase of a SU(2)  gauge theory.
Despite of the bias towards the trivial centre sector, we have
$p_0<1$ because of centre sector tunneling. The crucial observation
is that while increasing the chemical potential $\mu $ to approach the
mass $m$ from below, $\rho _1$ 
develops a singularity
while $\rho_0$ is perfectly finite. This implies that the weights
are given by $w_0=0$, $w_1=1$ for $\mu \to m$. Thus, $\rho (E,T,\mu)$
is approximately given by
$$
\rho (E,T,\mu) \approx - \frac{1}{ \mathrm{e}^{[E - \mu]/T }
- 1 }
$$
and, hence, signals the BEC instability for $\mu \to m$.
Note that the sign of $\rho $ is dictated by the centre element
$z_1=-1$. Because of this sign, the contribution to the baryon number
density is negative and arises from centre dressed quarks.
This contribution is genuinely different
from the contribution from bare anti-quarks which can be neglected
for the present choice of parameters. \vspace{.3cm}
\hfill \break
In conclusions, we have shown that the gluonic average of the 
quark determinant of $SU(2N)$
gauge theories does not depend on the type of boundary conditions for
the quark fields as long as the centre symmetry is realised.
Lattice gauge simulations for the gauge groups $SU(2 \ldots 5)$
corroborate these findings. Our results supplement the recent findings
of~\cite{Bilgici:2009jy} where SU(2) deconfinement was brought in line
with the sensitivity of the quark determinant to the boundary conditions.
Our results may have far reaching phenomenological implications:
(i) they exclude the (pre)formation of a
quark Fermi sphere at finite densities in the confinement phase;
(ii) since quarks are dressed with centre fields of the gluonic background
they escape the spin-statistics connection. This is in analogy to the anyons
in solid state physics~\cite{anyons}. While anyons only occur in 2+1
dimensions, our model is the first one of its kind which evades the
spin-statistics connection in 3+1 dimensions. (iii) Exotic states of
matter, such
as a Bose-Einstein condensate of quarks, might exist prior to
deconfinement induced by density.

\vspace{0.3cm}
\hfill \break
\noindent {\bf Acknowledgments:} Helpful discussions with G.~Dunne, H.~Gies,
K.Ya.~Glozman and M.~Rho are gratefully acknowledged.
This work was supported in parts by STFC under contract ST/H008853/1 
and by GRK 1523.

\end{document}